# Nondifractive propagation of light in photonic crystals

Kestutis Staliunas[1] and Ramon Herrero[2]

[1] *ICREA, Departament de Fisica i Enginyeria Nuclear, Universitat Politecnica de Catalunya,*

*Colom 11, E-08222 Terrassa, Barcelona, Spain.*

[2] *Departament de Fisica i Enginyeria Nuclear, Universitat Politecnica de Catalunya,*

*Comte Urgell 187, E-08036 Barcelona, Spain*

**Abstract:**

We show that diffraction of electromagnetic radiation (in particular of a visible light) can disappear in propagation through materials with periodically in space modulated refraction index, i.e. photonic crystals. In this way the light beams of arbitrary width can propagate without diffractive broadening and, equivalently, arbitrary light patterns can propagate without diffractive "smearing".

PACS numbers: 42.55.Tv, 42.25.Fx, 42.79.-e

Photonic crystals, i.e. the materials with refraction index periodically modulated in space on a wavelength scale [1], are well known to modify the properties of light propagation. Photonic crystals can introduce photonic band-gaps in radiation spectra, i.e. they can serve as light conductors or insulators [2]. Photonic crystals can modify the dispersion of the light, i.e. they can significantly reduce its group velocity [3]. We show in the letter that photonic crystals can also modify the diffraction of the light in that the diffraction can vanish in propagation through particularly prepared photonic crystals.

Diffractive broadening of light beams (and of wave envelopes in general) is a fundamental phenomenon limiting the performance of many linear and nonlinear optical devices. Geometrical interpretation of light- (and in general of wave-) diffraction is as follows: light beams of arbitrary shape can be Fourier decomposed into plane waves, which in propagation acquire phase shifts depending on their propagation angles. This dephasing of the plane wave components results in a diffractive broadening of the light beam. Fig.1.a illustrates normal diffraction in propagation through homogeneous material, where the longitudinal component of the wave vector depends trivially on the propagation angle: $k_{II} = k_z = \sqrt{|\mathbf{k}|^2 - |\mathbf{k}_\perp|^2}$, where $\mathbf{k}_\perp = (k_x, k_y)$. In general the normal, or positive diffraction means that the surfaces of the constant frequency are concave in the wave-vector domain $\mathbf{k} = (k_x, k_y, k_z)$, as illustrated in Fig.1.a.

It is known, that diffraction can become negative for materials with refractive index modulated in a direction perpendicular to the propagation direction. The negative diffraction, as illustrated in Fig.1.b, geometrically means that the surfaces of constant frequency are convex in wave-vector domain. The negative diffraction was predicted for the waves of electromagnetic radiation [4], for acoustic waves [5], and for matter waves [6]. The change of sign of the diffraction is extremely interesting for propagation in nonlinear materials, as resulting in a change between nonlinear self-focusing and defocusing.

We show in this letter that in propagation through materials with refractive index periodically modulated in two or three spatial dimensions (in two- or three-dimensional photonic crystals) diffraction can vanish to zero for particular frequencies of electromagnetic radiation, and for particular propagation angles. The expected phenomenon is illustrated in Fig.1.c, where the zero diffraction is supposed to occur in a particular point in the wave-vector domain where the curvature of the surfaces of constant frequency becomes exactly zero. Zero diffraction physically means that light beams of arbitrary width can propagate without diffractive broadening and, equivalently, that arbitrary light structures can propagate without diffractive "smearing".

Nondiffractive light propagation can be expected in differently fabricated photonic crystals with different symmetries of the lattices, and with different shapes of inclusions of enhanced (or reduced) refraction index. In this letter we restrict to a simplest case of a harmonic photonic crystal – a material with sinusoidal modulation of refraction index in two- and three- dimensional space. Such harmonic photonic crystal could be produced e.g. holographically, i.e. by light interference on a photorefractive material [7]. Interference of two (or four-) pairs of laser beams could result in two- (or three-) dimensional dynamical photonic crystals as illustrated in Fig.2. In the present letter we concentrate on a two-dimensional case, however simulations show that the phenomena persist in case of three-dimensions.

We consider a superposition of two periodic lamellae-like refraction index gratings: $\Delta n(\mathbf{r}) = 2m(\cos(\mathbf{q}_1\mathbf{r}) + \cos(\mathbf{q}_2\mathbf{r}))$ with $|\mathbf{q}_1| = |\mathbf{q}_2| = q$ at angles $\pm a$ to the optical axis, as shown in Fig.2. This results in refractive index modulation profile $\Delta n(x,z) = 4m\cos(q_\perp x)\cos(q_{II} z)$, with $q_{II} = q\cos(a)$, and $q_\perp = q\sin(a)$. The crystallographic axes of such a harmonic photonic crystal are $\mathbf{p} \cdot (\pm 1/q_\perp, 1/q_{II})$, and the reciprocal lattice vectors of the photonic crystal are $\mathbf{q}_1$ and $\mathbf{q}_2$. We further assume for simplicity that the spatial period of photonic crystal is significantly larger than the wavelength of the probe beam [8]. This legitimates paraxial approximation for the description of propagation of the probe beams:

$$(2ik_0 \, \partial/\partial z + \partial^2/\partial x^2 + 2\Delta n(x,z)k_0^2)A(x,z) = 0. \qquad (1)$$

Here $A(x,z)$ is the slowly varying complex envelope of the electromagnetic field in two-dimensional space $E(x,z,t) = A(x,z)e^{ik_0 z - iw_0 t}$ propagating along z-direction (along the diagonal direction of the rhombs in the two-dimensional case of Fig.2) with the wave-number $k_0 = w_0/c$. First we perform an analytical study of the propagation of the plane waves, by expanding the electromagnetic field into a set of spatially harmonic (Bloch) modes. Then we prove the nondiffractive propagation by a direct numerical integration of wave equation (1).

The technique of our harmonic expansion is analogous to expansion in finite set of Bloch modes, as e.g. described in [9]. We include 5 most relevant harmonics in the expansion (respectively 9 modes for three-dimensional case):

$$A(x,z) = e^{ik_\perp x + ik_{II} z}\left(A_0 + A_1 e^{+iq_\perp x + iq_{II} z} + A_2 e^{+iq_\perp x - iq_{II} z} + A_3 e^{-iq_\perp x + iq_{II} z} + A_4 e^{-iq_\perp x - iq_{II} z}\right). \quad (2)$$

consisting of one central component with the wave vector $\mathbf{k} = (k_\perp, k_{II})$, and four most relevant modulated (sideband) components with the wave-vectors $\mathbf{k}_i = (k_{\perp,i}, k_{II,i}) = (k_\perp \pm q_\perp, k_{II} \pm q_{II})$ respectively. This results in a system of coupled linear equations for the complex amplitudes of the harmonic components:

$$\left(-2k_0 k_{II} - k_\perp^2\right)A_0 + 2mk_0^2(A_1 + A_2 + A_3 + A_4) = 0 \quad (3.a)$$

$$\left(-2k_0 k_{II,i} - k_{\perp,i}^2\right)A_i + 2mk_0^2 A_0 = 0 \quad , \quad i = 1,2,3,4 \quad (3.b)$$

The transversal dispersion relation (the dependence of the longitudinal component $k_{II}$ on the transverse component $k_\perp$ of the wave-vector) can be calculated from (3), however the expression is too complicated for analytical interpretation. The numerical plots of transverse dispersion relation are given in Fig.3. In the absence of refractive index modulation $m = 0$ the formal solution of (3) consist of a set of parabolas (dashed curves in Fig.3.a) shifted by the vectors of photonic crystal lattice $\mathbf{q}_{1,2}$ one with respect to another. These parabolas represent the transverse dispersion curves for uncoupled harmonic components of the expansion (1). We note that in nonparaxial description of the light propagation these parabolas would be substituted by circles. The modulation of the refractive index $m \neq 0$ lifts the degeneracy at the crossing points and gives rise to band-gaps in spatial wave-number domain (Fig.3.a). We focus however in this letter not in the appearance of the band-gaps, but rather on appearance of platoes of transverse dispersion curves, indicating the vanishing of diffraction. Numerical solution of (3) yields that the curvature of the spatial dispersion curve become exactly zero at some set of points in wave-vector domain, corresponding for a particular choice of parameters (essentially the geometry of photonic crystal given by vectors $\mathbf{q}_{1,2}$ and modulation depth m) as shown in Fig.3.b. Among the five dispersion branches, the upper one can become nondifractive, with zero curvature at $k_\perp = 0$. The insets of Fig.3 show the field eigenfunctions at a particular propagation distance (the envelopes $A(x, z = 0)$). The nondifractive dispersion branch corresponds to the most homogeneous Bloch mode as shown by the insets to Fig.3.

Next we use adimensional variables: the transverse wave-number of light is normalized to the wave-number of transverse modulation of refraction index, $K_\perp = k_\perp / q_\perp$, and the longitudinal wave-numbers are normalized by $(K_{II}, Q_{II}) = (k_{II}, q_{II}) \cdot 2k_0 / q_\perp^2$. The space coordinates are thus rescaled as $Z = z q_\perp^2 / 2k_0$ and $(X, Y) = (x, y) q_\perp$. Two significant parameters remain after the normalization: $f = 4m k_0^2 / q_\perp^2$ which represents the modulation depth of the plane waves propagating in the photonic crystal, and $Q_{II} = 2 q_{II} k_0 / q_\perp^2$ which is proportional to the angle between the crystallographic axes of the photonic crystal.

Diffraction coefficient, i.e. a curvature of the transverse dispersion curve $d_2 = 1/2 \cdot \partial^2 K_{II} / \partial K_\perp^2$, becomes zero for a particular relation between parameter values of the photonic crystal. Fig.4 shows the zero diffraction regimes in the space of normalized parameters $(f, Q_{II})$. The purely nondifractive regime can be obtained on a curve, i.e. on the area of a measure one in the two-dimensional parameter space. The strength of the effect, which is proportional to the size of the nondifractive area in the spatial wave-number domain, depends strongly on the location along the zero diffraction line, as insets of Fig.4 indicate.

An analytical form of the zero diffraction curve is in general complicated. A simple asymptotical expression can be, however, found in a limit of weak refractive index modulation $f \ll 1$, when only three harmonics of the harmonic expansion (2) are relevant: the homogeneous one, and those with $q_{II} < 0$ (three upper parabolas in Fig.3.a). The power series expansion of the transverse dispersion relation in this limit results in $Q_{II}|_{d_2=0} \approx 1-2f^{2/3}$, allowing a simple relation between the scaled variables, and matching with the numerically calculated curve in Fig.4 in the limit $f \ll 1$.

We checked the above predicted phenomenon of nondifractive propagation by direct numerical integration of the equation for the light wave propagation (1) for two- and three-dimensional modulation of refractive index. The results for two-dimensional case are shown in Fig.5. The integration results in a evidence of the suppression of the diffractive spreading of a narrow probe beam: whereas in the absence of spatial modulation of the refractive index the light beam is diffractively broadening (Fig.5.a), in the presence of refraction index grating of particular parameters the spreading was strongly suppressed (Fig.5.b). The used set of parameters $(f,Q_{//})=(0.28, 0.57)$ lies close to analytically calculated zero dispersion curve of Fig.4, and are chosen in order to obtain a diffraction minimum for the Gaussian envelope [10]. The magnified part of nondifractive propagation plot (Fig.5.e) indicates that the nondifractively propagating beam is in fact an envelope of the spatially modulated nondifractive propagation mode. We note that our calculations in three-dimensional case result in essentially the same nondifractive behavior as in two-dimensional cases shown in Fig.5.

As diffraction (the curvature of the dispersion curve) becomes exactly zero on a curve in wave-number domain then, strictly speaking, nondifractive propagation has a sense for infinitely broad beams only. Spatial Fourier spectrum of beams of finite width occupies an extended area in a wave-vector space (indicated area in Fig.1.c), where the dephasing of spatial Fourier components, being not strictly zero, remains negligibly small for a finite propagation distance. The minimal transverse size of a light structure $\Delta x$ which can propagate with a negligible diffractive broadening over a given distance L is inversely proportional to the width of the area of negligible diffraction in wave-number domain $\Delta k_\perp$: $\Delta x \approx 2/\Delta k_\perp$. The distance of nondifractive propagation L is related with the size of nondifractive area by $dk_{II} L \propto 1$, where $dk_{II}$ indicates the small variation of the longitudinal component of the wave-vector within the nondifractive area. We analyzed this weakly-difractive propagation by series expansion of the transverse dispersion curve around $K_\perp = 0$: $K_{II}(K_\perp) = \sum_{n=0}^{\infty} d_n K_\perp^n$, $d_n = 1/n! \cdot \partial^n K_{II}/\partial K_\perp^n$. As the second order derivative is zero at zero diffraction point, and all odd order derivatives are zero at $K_\perp = 0$ because of symmetry, then the fourth order derivative (fourth order diffraction) is the lowest and most relevant one. Approximation of the dispersion curve by the fourth order power function leads to the following evolution equation:

$$\left(\partial/\partial Z + d_4 \partial^4/\partial X^4\right) A_B(X,Z) = 0. \tag{4}$$

for the spatial envelope of the nondifractive Bloch mode $A_B(X,Z)$. $d_4$ is an adimensional coefficient for the fourth order diffraction of order of magnitude of one.

The analysis of the propagation of the light beam obeying (4) is straightforward in spatial Fourier domain:

$$A_B(K_\perp, Z) = A_B(K_\perp, Z=0) \cdot \exp\left(id_4 K_\perp^4 Z\right) \tag{5}$$

resulting, however, in not a trivial propagation picture in the space domain. Considering an infinitely narrow initial beam ($\delta$-beam), with a flat spatial spectrum, equation (5) yields that the spatial components with $|K_\perp|^4 > |\Delta K_\perp|^4 \approx \pi/(d_4 Z)$ dephase significantly ($\Delta\varphi \geq \pi$) over a propagation distance Z. Assuming that the coherent (not dephasing) spatial spectrum components $|K_\perp| < |\Delta K_\perp|$ results in a nearly-Gaussian beam of half-width $\Delta X \approx 2/\Delta K_\perp$ the following expression for the evolution of the half-width of the beam is obtained: $\Delta X^4(Z) \approx 16 d_4 Z/\pi$, which in terms of the real world parameters reads:

$$\Delta x^4(z) \approx 8 d_4 z/\left(\pi k_0 q_\perp^2\right) \tag{6}$$

The evolution of nonzero width Gaussian beams can be directly calculated by convolution of the Gaussian envelope with the envelope of the diffracted $\delta$-beam given by (6). Fig.6.a shows the evolution of the half-width of the Gaussian beams of different initial half-width $\Delta x_0$ as calculated by solving numerically the propagation equation (4). For long propagation distances $z > \pi k_0 q_\perp^2 \Delta x_0^4/(8 d_4)$ the beam broadening follows the asymptotic fourth root dependence as evaluated above (6). Fig.6.b. shows the spatial envelope of the weakly-diffractively propagating beam. The central (weakly broadening) peak consists of in-phase $|K_\perp| < |\Delta K_\perp|$ spatial Fourier components, and the background radiation consists of dephased spatial Fourier component with $|K_\perp| > |\Delta K_\perp|$.

Concluding, we predict the nondifractive broadening of electromagnetic radiation in propagation through photonic crystals. We calculate the parameters of photonic crystal, and we evaluate the limits of nondifractive propagation. The limits of nondifractive propagation depend on the higher order (predominantly fourth order-) diffraction, i.e. depend on the position on the zero diffraction curve in parameter space. Assuming for a rough estimation that the spatial scale of refraction index modulation can be in principle reduced down to a half-wavelength of the propagating light $|q_\perp| \approx 2|k_0|$ (which is typical value for photonic crystals), and that the adimensional fourth order diffraction is of order of one, one obtains: $\Delta x^4(z) \approx zl^3/(4p^4)$. This gives a realistic estimation of the minimal width of the nondifractive beam to be propagated over distance z (or, equivalently the spatial resolution of the pattern to be transferred over the distance z). As a numerical example – the nondifractive propagation length of the beam of half-width of $10\ mm$ and $l = 1\ mm$ should be of the order of $1m$, whereas in diffractive propagation the corresponding Rayleigh length is about $0.3mm$.

The nondifractive propagation reported in the present letter is at the basis of the extremely small size cavity solitons (mid-band solitons) in nonlinear resonators with transversally modulated refraction index, as recently reported [11], but interpreted in a different way.

The work was financially supported by project FIS2004-02587 of the Spanish Ministry of Science and Technology. We gratefully acknowledge discussions with C.Cojocaru and J.Trull.

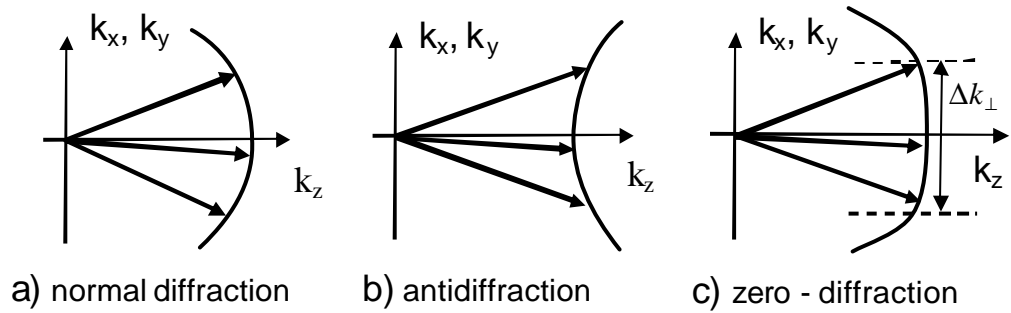

*Fig.1.* Geometrical interpretation of diffraction of electromagnetic radiation propagating along the z axis: a) positive, or normal diffraction in propagation through homogeneous materials; b) negative, or anomalous diffraction e.g. in propagation through spatially periodic structures; c) zero diffraction. The area of negligible diffraction is indicated.

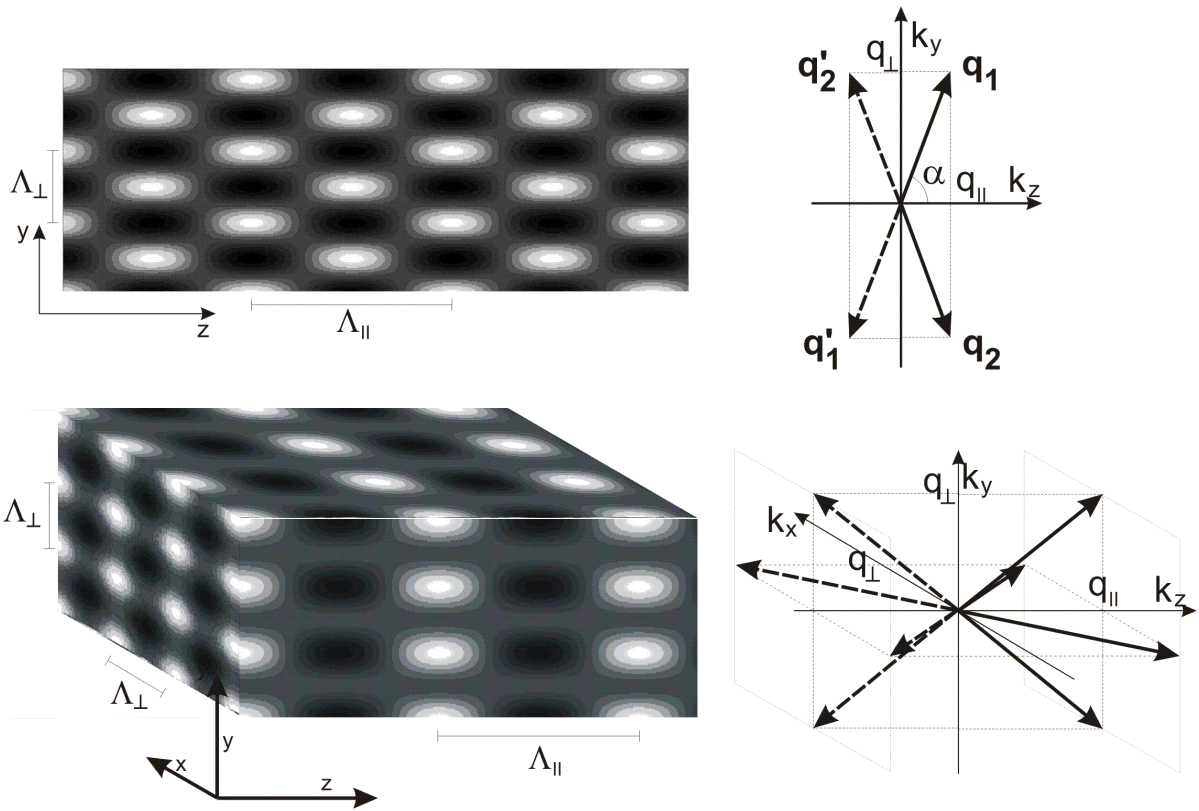

*Fig.2.* Harmonic modulation of refraction index as e.g. imposed holographically in photorefractive material. Upper (bottom) plots illustrate two- (three-) dimensional case, where grating is written by two (four) pairs of counterpropagating beams.

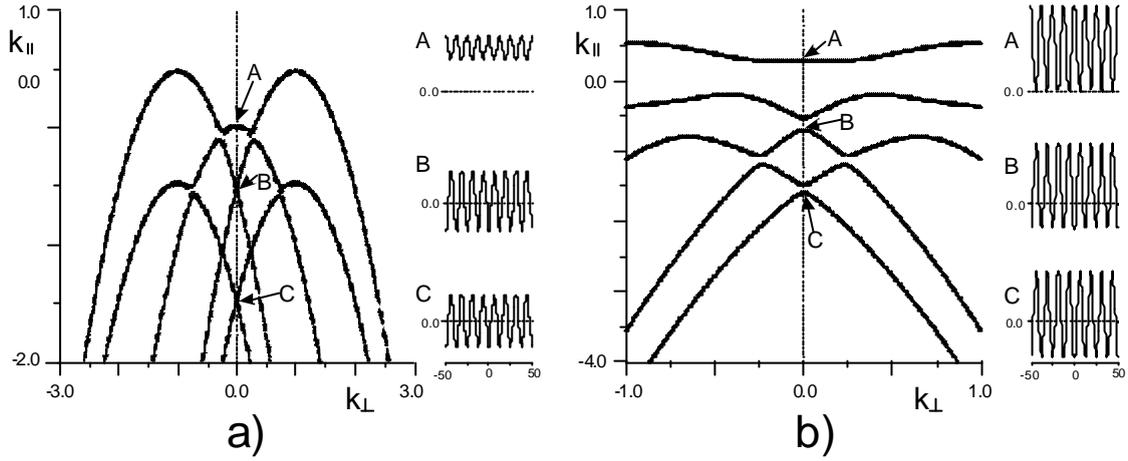

*Fig.3* Longitudinal component of wave-vector of Bloch modes $k_{\parallel}$ depending on the transverse component $k_{\perp}$, as obtained by numerical integration of (3): a) in absence of modulation m=0 (dashed parabolas), and in presence of weak modulation of refraction index $m = 0.003$ (solid line); b) for particular amplitude of modulation $m = 0.0175$ resulting in appearance of zero diffraction (zero curvature). Insets show corresponding eigenfunctions (Bloch modes) calculated at $k_\wedge$=0. Parameters used: $l = 500 nm$ ($k_0 = 4\pi\, 10^6 m^{-1}$), $q_{\perp} = 2\pi 10^6 m^{-1}$ and $q_{\parallel} = 0.75\, 10^6 m^{-1}$.

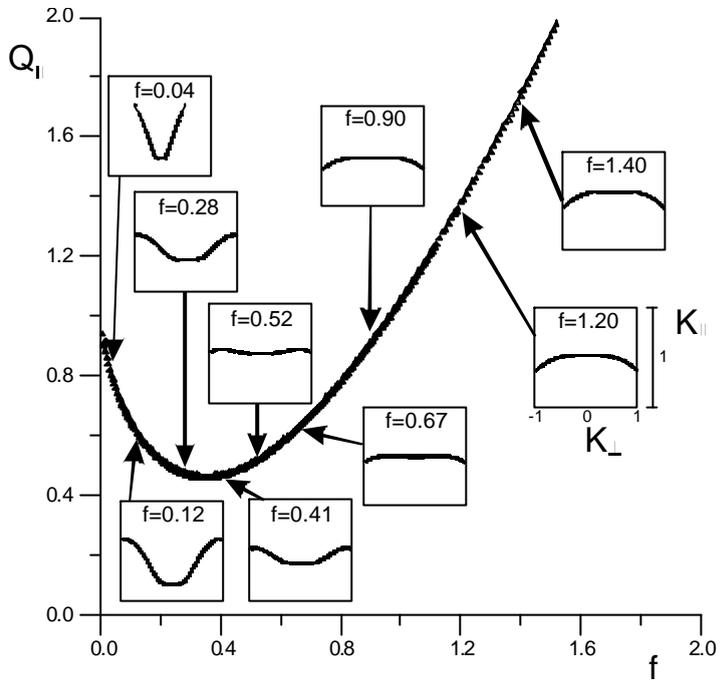

*Fig 4. Zero diffraction curve in space of adimensional parameters $(f, Q_{II})$, as obtained by numerical integration of (3). Insets correspond to the first branch of diagram $(K_\perp, K_{II})$ arround $K_\perp = 0$.*

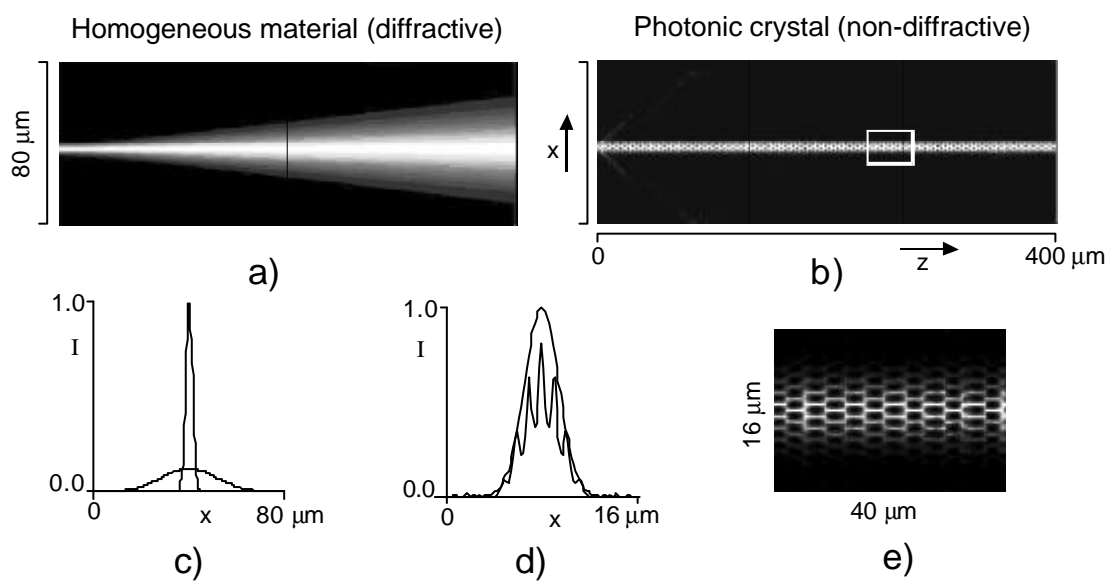

***Fig.5.*** *Diffractive (a) and nondifractive (b) propagation of a Gaussian beam, as obtained by numerical integration of (1) in two-dimensional case. The parameters used for calculation correspond to the point $(f, Q_{//})=(0.28, 0.57)$ in Fig.4. c) and d) show initial and final envelopes of the beam (intensity), and e) shows the magnified area from two-dimensional plot in b). The real world parameters: $\lambda=500$ nm, $n_0=1$, $m=0.0175$, $l_{0\perp}=1.0$ mm and $l_{0//}=7.0$ mm. The initial beam waist $W_0=2.82$ mm. with corresponding Rayleigh length: $z_0=50$ mm.*

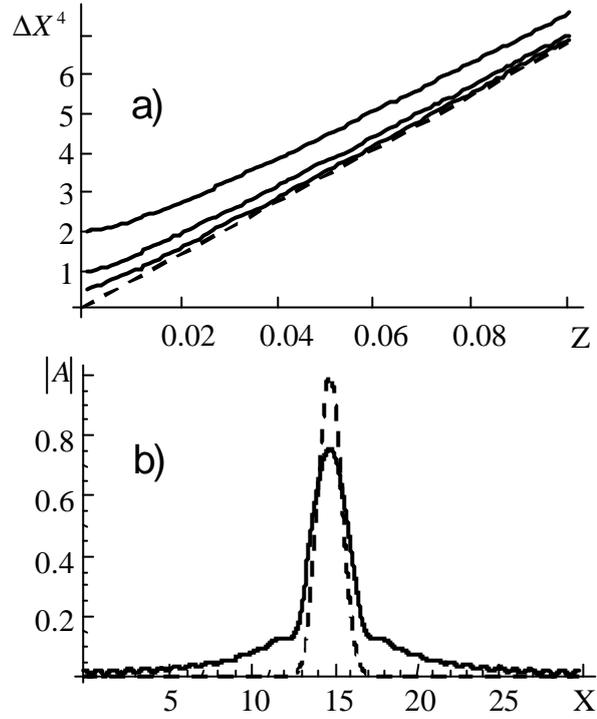

***Fig.6*** *Propagation of Gaussian beam in presence of fourth order diffraction, as obtained by numerical integration of (4) for two-dimensional case, with $d_4 = 1$: a) evolution of the half-width of the Gaussian beams for initial half-widths: $\Delta x_0^4 = 0.5, \ 1, \ and \ 2$. Note that the half-width in fourth power is depicted. Dashed line is for guiding the eye; b) spatial profiles of the initial $Z = 0$ (dashed) and the final $Z = 0.1$ beam. Note the broadening of the central part of the beam, and the appearance of the background radiation.*